\documentclass[journal, hideappendix]{vgtc}                     

\onlineid{1483}

\vgtccategory{Research}

\vgtcpapertype{Representation \& Interaction}

\title{SpreadLine: Visualizing Egocentric Dynamic Influence}

\author{%
  \authororcid{Yun-Hsin Kuo}{0009-0000-1891-8993},
    \authororcid{Dongyu Liu}{0000-0002-8915-2785}, and 
    \authororcid{Kwan-Liu Ma}{0000-0001-8086-0366}
}

\authorfooter{
  \item Yun-Hsin Kuo, Dongyu Liu, and Kwan-Liu Ma are with the University of California, Davis. E-mail: 
\{yskuo, dyuliu, klma\}@ucdavis.edu
}

\abstract{
Egocentric networks, often visualized as node-link diagrams, portray the complex relationship (link) dynamics between an entity (node) and others. However, common analytics tasks are multifaceted, encompassing interactions among four key aspects: strength, function, structure, and content.
Current node-link visualization designs may fall short, focusing narrowly on certain aspects and neglecting the holistic, dynamic nature of egocentric networks. 
To bridge this gap, we introduce SpreadLine, a novel visualization framework designed to enable the visual exploration of egocentric networks from these four aspects at the microscopic level. 
Leveraging the intuitive appeal of storyline visualizations, SpreadLine adopts a storyline-based design to represent entities and their evolving relationships. We further encode essential topological information in the layout and condense the contextual information in a metro map metaphor, allowing for a more engaging and effective way to explore temporal and attribute-based information.
To guide our work, with a thorough review of pertinent literature, we have distilled a task taxonomy that addresses the analytical needs specific to egocentric network exploration. 
Acknowledging the diverse analytical requirements of users, SpreadLine offers customizable encodings to enable users to tailor the framework for their tasks. 
We demonstrate the efficacy and general applicability of SpreadLine through three diverse real-world case studies (disease surveillance, social media trends, and academic career evolution) and a usability study. 
}


\graphicspath{{figs/}{figures/}{pictures/}{images/}{./}} 

\PassOptionsToPackage{svgnames}{xcolor}
\usepackage{colortbl}
\usepackage[most]{tcolorbox}

\usepackage{tabu}                      
\usepackage{booktabs}                  
\usepackage{lipsum}                    
\usepackage{mwe}                       

\usepackage{mathptmx}                  
\usepackage{soul}
\usepackage{pdfpages}
\usepackage{enumitem}
\usepackage{caption}
\usepackage{balance}

\begin{document}
\newcommand{\yunhsin}[1]{{\color{teal}[YHK\@: #1]}}
\newcommand{\revise}[1]{{\color{black}#1}}

\newcommand{\egonetwork}[0]{egocentric networks}
\newcommand{\entirenetwork}[0]{entire networks}

\newcommand{\structure}[0]{{\color{Structure} \textbf{structure}}}
\newcommand{\function}[0]{{\color{teal} \textbf{function}}}
\newcommand{\strength}[0]{{\color{Strength} \textbf{strength}}}
\newcommand{\content}[0]{{\color{Content} \textbf{content}}}

\definecolor{cback}{HTML}{EDEFF1}
\definecolor{cframe}{HTML}{B9C4CA}
\definecolor{cgrey}{HTML}{666666}
\definecolor{cT}{HTML}{E5D9F2} 
\definecolor{cF}{HTML}{bae6fd }
\definecolor{cR}{HTML}{D8F793}
\definecolor{BG}{HTML}{ffffff}

 \newtcbox{\inlineFbox}[2][]{enhanced,
 box align=base,
 nobeforeafter,
 colback=#2!15!BG,
 colframe=#2!50!BG,
 size=small,
 toprule=0.75pt,
 bottomrule=0.75pt,
 leftrule=0.75pt,
 rightrule=0.75pt,
 fontupper=\footnotesize,
 left=1pt,
 right=1pt,
 boxsep=1.2pt,
 #1}

\newcommand{\structureBox}[1]{{\inlineFbox{Structure}{\textbf{#1}}}}
\newcommand{\functionBox}[1]{{\inlineFbox{Function}{\textbf{#1}}}}
\newcommand{\strengthBox}[1]{{\inlineFbox{Strength}{\textbf{#1}}}}
\newcommand{\contentBox}[1]{{\inlineFbox{Content}{\textbf{#1}}}}

\definecolor{Structure}{HTML}{7fa603} 
\definecolor{Strength}{HTML}{7c258f}
\definecolor{Function}{HTML}{3ab0a7} 
\definecolor{Content}{HTML}{d97c56}

\definecolor{dypink}{HTML}{ec008c}
\newcommand{\dyu}[1]{{\color{dypink} #1}}
\newcommand{\dst}[1]{{\st{#1}}}
\newcommand{\rev}[2]{{\st{#1}\textcolor{dypink}{#2}}}

\definecolor{NavyBlue}{RGB}{0, 0, 185}


\firstsection{Introduction}

\maketitle

Networks, as a data structure, are commonly
used to represent interactions among entities.
Such a structure in real-world applications may contain complex and dynamic information: the entities and their ever-changing relations can be associated with a set of attributes that portray their characteristics and evolution~\cite{ahn2014task, kale2023state}.
The egocentric perspective in networks focuses on the relationship dynamics between an entity and others, which is crucial for analyzing individual entities.
Inherited from the nature of networks, egocentric networks are also challenged by the complexity of temporal multivariate information.
%

In egocentric network analysis, as suggested by Perry et al.~\cite{perry2018egocentric}, 
four aspects of the network: \strength{}, \function{}, \structure{}, and \content{} 
should be considered. 
Strength represents the intensity of edges, whereas function considers the types of edges.
Structure refers to the network topology, and finally, content describes the characteristics of entities.
For example, a social network contains humans as entities (or nodes) with their social relationships as edges (\structure). 
A behavior of the ego (\content) may be motivated by the entities connected as family members, friends, or coworkers (\function). 
These relations can be further specified by how long they have known each other (\strength).

Visual analytics tools have been developed to support diverse network analysis tasks~\cite{kale2023state}.
However, these tasks are often multifaceted, as the network effects are a function of interaction among the aforementioned four network aspects \cite{perry2018egocentric}.
\revise{
In the social network example, a nonsmoker may adopt the habit of smoking (content) due to active (strength) peer pressure (structure) from a group of friends (function)~\cite{simons2010recent}.
Node-link diagrams, a conventional representation 
of networks~\cite{archambault2014temporal, hadlak2015survey}, effectively present structure to reveal the scale of the influence. 
However, they may have limited capacity to present all other aspects at a time to meet diverse analysis needs, neglecting the holistic nature of egocentric networks, such as the direction of the influence.
This limitation motivates us to explore how to visualize a wider variety of information to understand egocentric networks coherently.
} 


Analyzing relationship dynamics between entities is similar to understanding interactions between a story's protagonist and other characters.
Storyline visualizations are effective at revealing the characters' behaviors in a story~\cite{ogawa2010software, tanahashi2012design}.
Each character represents one line, where their close alignment indicates their involvement in important story events.
The user can trace through a line of an entity to understand its dynamic behaviors and relationships with others as the story advances. 
This design rationale aligns with the foundation of many analysis tasks of dynamic multivariate networks~\cite{ahn2014task} and allows other marks and channels to encode more information;
we are therefore inspired to explore the visualization design of dynamic multivariate egocentric networks.

We introduce a novel visualization design framework that supports the exploration of dynamic multivariate networks from the egocentric perspective. 
To guide the framework development, we have conducted a thorough literature review and distilled a task taxonomy that addresses the analytical needs specific to individual egocentric networks (i.e., the microscopic level).
Building upon the storyline-based design, our visualization encodes the essential topological information 
in the layout.
Inspired by the aesthetics of metro maps~\cite{wu2020survey}, we condense the contextual information in the metaphor that helps the user identify the timestamp or entity of analysis interest to drill down for further details.
SpreadLine offers customizable encodings to fulfill the diverse exploration needs of users, for which we provide guidelines on configuring the framework.
We demonstrate the applicability of SpreadLine using three real-world datasets in different domains.
To evaluate the interpretability and effectiveness of SpreadLine, we recruit ten participants, with different levels of experience in visualization, and conduct a usability study.
The study results confirmed SpreadLine's ability to connect the four network aspects visually.

The main contributions of this work are: 
(1) Delineating a task taxonomy of egocentric network visual analysis at the microscopic level; 
(2) \revise{Introducing SpreadLine}, a flexible visualization framework for comprehensive egocentric network exploration; and
(3) Demonstrating the generalizability and effectiveness of SpreadLine through three diverse real-world case studies  and a usability study.

\section{Related Work}

\subsection{Egocentric Network Visual Analysis}
\label{sec:egocentricl-visual-analysis}
We describe essential terms in egocentric networks and then examine how relevant works support analysis tasks.
\revise{Often in major literature\cite{prell2011social}, egocentric networks contain only the ego and its 1-level alters (i.e., nodes directly connected to the ego).} 
Some research extends to more distant connections, e.g., 2-level alters (nodes directly connected to ego's 1-level alters)~\cite{zhao2016egoline}.
For simplicity, we address \structure{}, \strength{}, \function{}, and \content{}
as ``the four network aspects''.

\begin{table*}[t]
    \small
    \centering
    \includegraphics[width=1.0\textwidth]{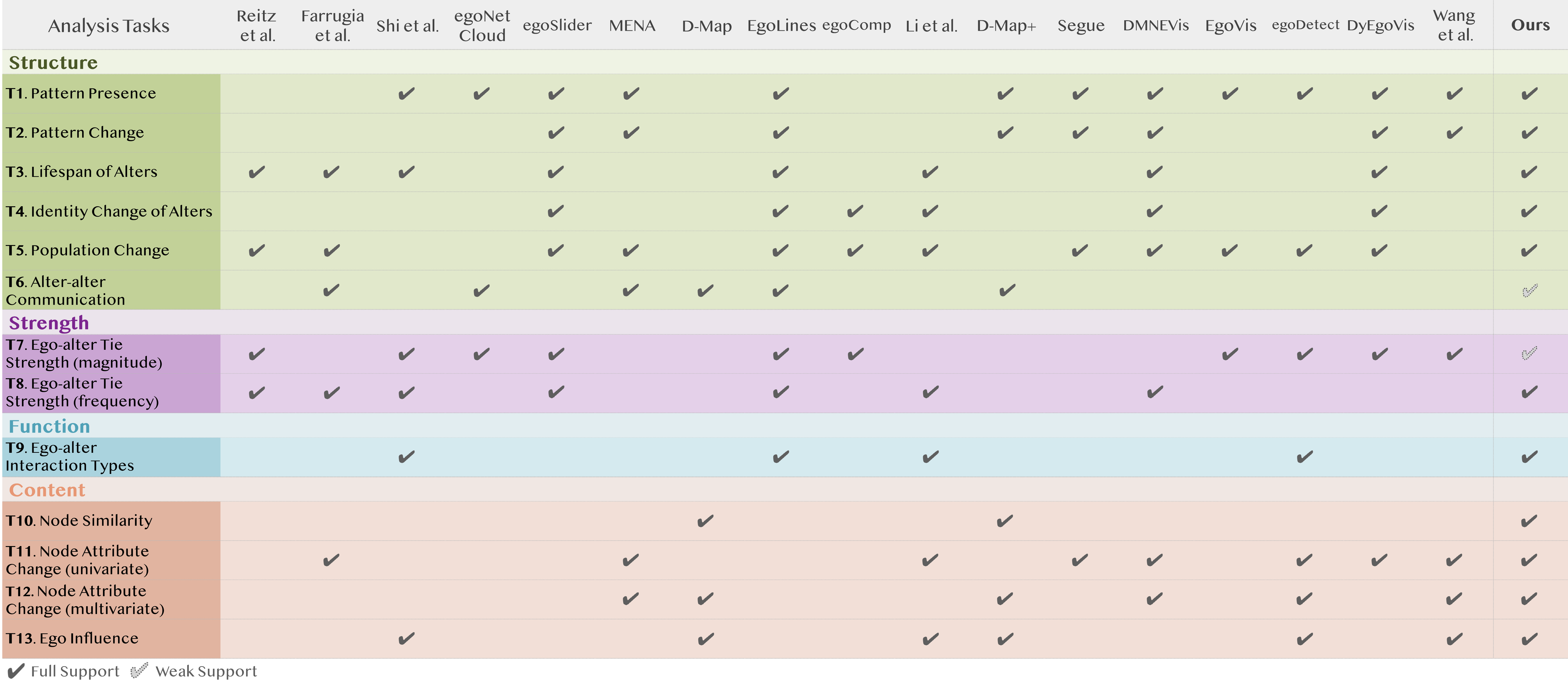}
    \caption{Task taxonomy for egocentric network analysis at the microscopic level. We include visualization frameworks~\cite{reitz2010framework, shi20141, zhao2016egoline, wang2022egocentric} and visual analytics tools~\cite{farrugia2011exploring, liu2015egonetcloud, wu2015egoslider, he2016mena, chen2016d, chen2018d, liu2017egocomp, li2017visual, law2018segue, peng2018dmnevis, lu2020egovis, pu2020egodetect, fu2021dyegovis} for egocentric networks. Our taxonomy is derived from surveys~\cite{ahn2014task, kale2023state}, a textbook~\cite{perry2018egocentric}, and the analysis tasks listed in relevant works~\cite{wu2015egoslider, he2016mena, zhao2016egoline, chen2016d, shi20141}, where we generalize our task descriptions to only concern network properties.}
    \label{fig:taxonomy}
    \vspace{-0.2in}
\end{table*}

We consider visualization frameworks or visual analytics tools designed for egocentric networks as pertinent to our work.
To provide a systematic examination of egocentric network visual analysis, we summarize an initial list of common analysis tasks by integrating the task taxonomy developed in other works~\cite{wu2015egoslider, zhao2016egoline, he2016mena, fu2021dyegovis, chen2018d, ahn2014task, shi20141}.
As many works are designed for various purposes (e.g., global network -- egocentric network -- individual node~\cite{wu2015egoslider}) or application scenarios (e.g., anomaly detection~\cite{pu2020egodetect}), we generalize the tasks as descriptions of network properties and reduce the list to concern only egocentric networks.
With the new taxonomy, we examine whether relevant works support the tasks through encoding or interaction design of egocentric network visualization and auxiliary views. 
Our taxonomy is categorized by the four network aspects, as shown in~\autoref{fig:taxonomy}.
%
Following the analysis tasks in order, we clarify the differences between some tasks.
For alter identity, \structureBox{T4} focuses on structural changes, such as \revise{alters leaving and rejoining the network as 1-level or 2-level alter}, whereas \contentBox{T11} includes the contextual identity change of alters (e.g., friend identity in social networks).
As Perry et al.~\cite{perry2018egocentric} emphasized that \strength{} has two aspects (magnitude and frequency), also evident in the analysis tasks mentioned in~\cite{shi20141}, we made a distinction of \strengthBox{T7} and \strengthBox{T8}.
We differentiate the tasks of node attribute change (\contentBox{T11} and \contentBox{T12}) to highlight the capability of handling multivariate information.
Finally, we illustrate what node attributes are considered \content{} instead of \structure{}. 
We exclude node attributes derived from the network topology if they are inherently considered in other analysis tasks. 
For example, ``the number of alters'' is considered as \structureBox{T5}, or ``the average ego-alter tie strength'' concerns \strengthBox{T7}.
While ``centrality'' is computed from network topology, as it refers to a node's influence, we deem it related to \contentBox{T13}.
We adopt common definitions from other taxonomies for the remaining tasks: \structureBox{T1}, \structureBox{T2}, \structureBox{T3}, \structureBox{T6}, \functionBox{T9}, and \contentBox{T10}.


\revise{In \autoref{fig:taxonomy}, we observe that many works, where a majority adopts node-link representations~\cite{reitz2010framework, farrugia2011exploring, liu2015egonetcloud, he2016mena, liu2017egocomp, lu2020egovis}, only support two of the four network aspects} (e.g., \structure{} paired with \strength{} or \content{}), with few extending to three. 
Some works covered four network aspects~\cite{shi20141, li2017visual, pu2020egodetect}, but two of them were designed for specific applications~\cite{li2017visual, pu2020egodetect}. 
The difference between our framework and Shi et al.~\cite{shi20141} lies in the visual representation and the design focus. They adopted a double-sided timeline of the ego in a radial layout to emphasize the ingoing and outgoing influence of the ego.
Our work is unique in that our framework 
flexibly and effectively supports all these tasks at the user demand for a comprehensive 
exploration of egocentric networks.




\subsection{Dynamic Egocentric Network Visualization}
\revise{Animated transitions and timeline-based methods are two main visualization techniques for dynamic networks~\cite{beck2017taxonomy}. To our best knowledge, many egocentric network visualizations detail the structural changes of a network on a timeline, such as small multiples~\cite{he2016mena}. While visualization techniques for different networks are also designed to present complex information~\cite{fischer2021towards, xu2024imvis, pena2022hyperstorylines}, here, we focus on only the visual representation specifically designed for egocentric networks.}

Node-link representation, while intuitively informing the network structure, is prone to visual clutter due to scalability.
To address this, existing works condensed essential network structure in the visual representation through extraction algorithms~\cite{liu2015egonetcloud} or user-defined groups~\cite{he2016mena, lu2020egovis}.
\revise{Radial layouts are also favored for alternative channels (radius and angle)}.
While angle was often employed to position alters, radius was used to represent the timeline~\cite{farrugia2011exploring}, the interaction duration between the ego and alters~\cite{reitz2010framework, shi20141}, or the alter identity~\cite{liu2017egocomp}. 
Existing works have explored other representations: hexagon maps~\cite{chen2016d, chen2018d} or visual metaphors  
 (``pollen spread''~\cite{wang2022egocentric} or trees~\cite{sallaberry2016contact, fung2016design}) to show ego influence on different alter communities, or stream graphs to characterize the population change by the alter identity (univariate \content{})~\cite{law2018segue}.

Line-based representation is the most relevant design to our work.
Unlike line marks in node-link representations that denote the relationship between entities, they connect the same entity across timestamps to facilitate visual tracking of an entity.
Several works designed additional encodings to highlight various network information.
EgoSlider~\cite{wu2015egoslider} used vertical space and contours to inform the identity change of an alter (\structure{}), where a similar approach was adopted to show the identity of alters (\content{})~\cite{peng2018dmnevis}.
EgoLines~\cite{zhao2016egoline} embedded a subway visual metaphor in lines to denote the network connectivity (\structure{}). 
DyEgoVis~\cite{fu2021dyegovis} used auxiliary lines to highlight shared alters across networks at the same timestamp (\structure{}).
A fundamental difference of our work is that our design incorporates the station style in the metro maps to encode the identity of alters (univariate \content{} or \structure{}) and considers \function{} and \strength{} during the layout generation. 
\vspace{-1pt}
\subsection{Storyline Visualization}
\label{sec:related-storyline-vis}
Inspired by Munroe's Movie Narrative Charts~\cite{xkcd}, Ogawa and Ma~\cite{ogawa2010software} introduced storyline visualization with layout heuristics to display interactions between software developers. 
A set of design principles was later refined, where three metrics --- line wiggles, line crossings, and white spaces --- were defined to measure layout quality~\cite{tanahashi2012design}.
StoryFlow~\cite{liu2013storyflow} formulated layout computation as a three-stage optimization problem: ordering, alignment, and compact, enabling layout generation at interactive speed.
While some works optimized their layout generation on quality metrics~\cite{dobler2023crossing, gronemann2016crossing, van2018crossing, kostitsyna2015crossing, froschl2017wiggles} or streaming data~\cite{tanahashi2015efficient}, other works employed storyline visualization in different domains~\cite{lu2014group, ogawa2010software, di2020actor, kim2010genealogy, yagi2015weather, balint2016eye, baumgartl2020search}, including dynamic social networks~\cite{reda2011storynetwork, arendt2014sven}.

There has been increasing attention to integrating various information into storyline visualizations, such as background contours to inform locations~\cite{liu2013storyflow, tanahashi2012design}.
iStoryline~\cite{tang2018istoryline} and PlotThread~\cite{tang2020plotthread} incorporated user specifications, e.g., aligning certain line segments, in authoring the storylines. 
Line styles~\cite{padia2019system} or line branches~\cite{di2020actor} are utilized to inform concurrent relationships of a character.
GeoStoryline~\cite{hulstein2022geo} introduced a design space on embedding geospatial maps in storyline visualizations.
The works mentioned above either tailored their layout generation algorithms or utilized visual encodings to inform additional information in their designs. 
We focus on relevant works that utilize the axes. 
SVEN~\cite{arendt2014sven} formulated ordering as a seriation problem to minimize line crossings and employed arc segments to depict the dynamic relationship among entities (\structure{}).
Others explicitly mapped information of the entities (univariate \content{}), onto the $y$-axis~\cite{yagi2015weather, reda2011storynetwork, arendt2017matters, baumgartl2020search, qiang2017storytelling}.
Our framework utilizes the $y$-axis to encode \function{}{} and \strength{}{} information by imposing constraints on the layout generation.


\vspace{-0.05in}
\section{Motivating Scenario}
\label{sec:scenario}
We provide a real-world example to argue the importance of combining information from the four network aspects.
From our other project collaboration~\cite{kuo2023investigating}, domain experts in animal surveillance intend to understand how various factors influence a disease outbreak at an animal farm.
We include technical network terms to illustrate the generalizability of our framework in other application scenarios.

\begin{figure}
    \centering
    \includegraphics[width=0.95\linewidth]{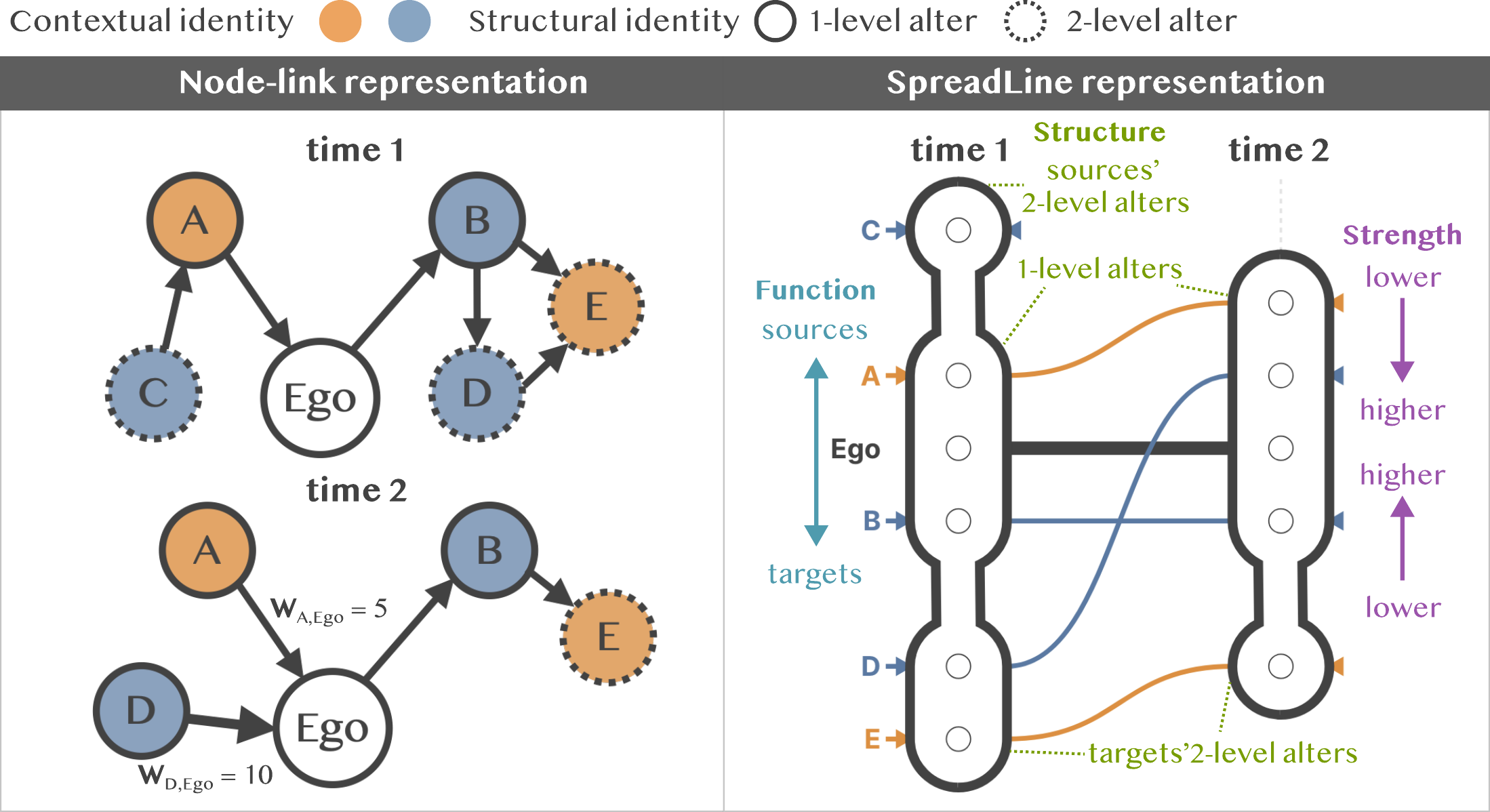}
    \caption{A foundational configuration of SpreadLine. Left: The node-link representation of the trade network among 6 animal farms across two timepoints. Right: Using the same information depicted on the left, SpreadLine employs various encodings to condense crucial network information, emphasizing the \function{}, \structure{}, and \strength{} aspects.}
    \label{fig:encodings}
    \vspace{-0.2in}
\end{figure}

In a community of animal farms (\textit{nodes}), the farms often communicate with each other through animal transport (\textit{directional edges}) due to their different roles (\content{}) in a production system.
\autoref{fig:encodings}-left presents an example illustration.
When an infectious disease outbreak is detected on a farm (\textit{ego}), the disease may propagate through close contact (e.g., transport, \structure{}) or geographical proximity (\content{}) to affect other farms, endangering the maintenance of animal health and production efficiency. 
Consequently, understanding the influence of the disease outbreak is crucial for animal health experts to develop strategies for disease prevention and control.
To better understand a disease outbreak, several analysis tasks are required:
(1) identifying farms (\textit{alters}) that either had interactions with the outbreak farm (\textit{egocentric network construction}) or located geographically close to the outbreak farm (\structure{} and \content{}); 
(2) narrowing down the farms by examining their health conditions (\textit{node attribute}) over time (\content{}); 
(3) constructing the potential disease propagation based on how farms interact with the outbreak farm (\textit{edge direction}, \function{}); and 
(4) quantifying the outbreak influence with the number of animals (\textit{edge weight}) involved in the risky animal movements (\strength{}).

Through these tasks, the experts can identify a list of risky farms for further analysis.
For example, farms that interact with a risky farm (\textit{alter-alter communication}) require extra attention for disease control. 
Experts can also investigate non-risky farms that were in contact with the outbreak farm to develop effective strategies for disease prevention.
We argue that these insights are challenging to obtain, had it not been for these tasks that consider information from the four network aspects.
\section{Designing SpreadLine}

In the forthcoming sections, we describe how SpreadLine translates essential network information from the four network aspects into coherent visual representations.
SpreadLine offers customizable encodings, allowing users to tailor the framework to their diverse analytical requirements.
We first present a foundational configuration of SpreadLine, followed by two more configurations in \autoref{sec:evaluation}. 
With the real-world example from \autoref{sec:scenario}, we illustrate how our encoding designs communicate complex information visually.


\begin{figure*}
    \centering
    \includegraphics[width=\textwidth]{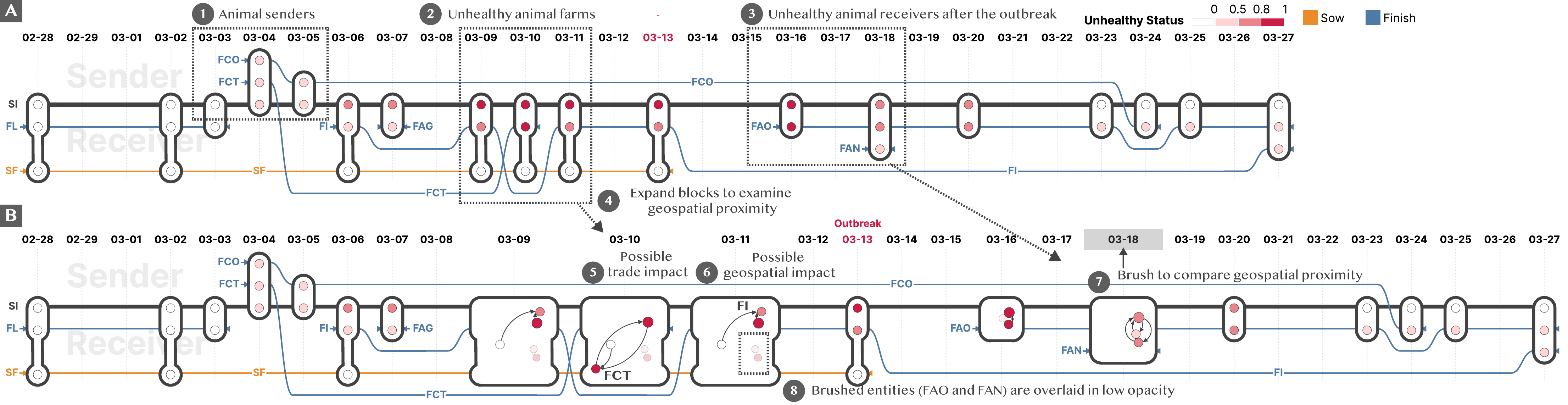}
    \caption{A disease outbreak detected on Farm SI (ego) on March 13th. To identify farms at risk of infection, an animal health specialist can investigate timepoints of interest (1, 2, and 3) and examine the farms from trading relationships, health conditions, and geospatial proximity (5, 6, and 8) through user interactions (4 and 7). \revise{The ego is indicated as a larger point in the expanded view. The layout is computed with \textbf{vertical space optimization}.}}
    \label{fig:animal}
    \vspace{-0.2in}
\end{figure*}

\vspace{-0.05in}
\subsection{Visual Design}
With a parallel comparison of node-link and SpreadLine representations of the same network in \autoref{fig:encodings}, we outline the high-level design of SpreadLine, followed by a discussion on the four network aspects.
SpreadLine uses continuous lines to portray each entity across timepoints, inspired by storyline visualization. 
Following timeline-based network visualization conventions, SpreadLine annotates timestamps and utilizes points to denote the entity presence at each timepoint.
These techniques together underscore SpreadLine's capacity to convey the temporal dynamics of the egocentric network (\structureBox{T1} and \structureBox{T2}).
Emphasizing the ego and its relationships with other entities (\contentBox{T13}), a core design of SpreadLine is to ensure the line representing the ego (referred to as ``the ego line'') remains consistently straightened and thickened.

These fundamental design rationales enable us to track a specific entity over time (\structureBox{T3}).
In the case of \autoref{sec:scenario}, farm tracking is vital for fulfilling those analysis tasks.
The remaining challenge lies in the visual translation of network information at each timepoint.
Our main approach involves vertically aligning the network population (\structure{}), utilizing space to characterize roles of the entities to the ego (\function{}), and assigning more important entities closer to the ego (\strength{}).
We adopt a drill-down approach by presenting the \content{} information in auxiliary views.
As depicted in \autoref{fig:encodings}, in this trade network example, \revise{we differentiate between 1-level and 2-level alters} with visual cues (e.g., nodes A and C), position direct animal senders and receivers differently (e.g., nodes A and B), and prioritize those that transported more animals with the ego (e.g., node D at time 2).

\textbf{Usage Scenario.}
To illustrate how these design rationales facilitate disease outbreak investigations, we consider the following scenario.
Suppose Jane, an animal health specialist, would like to investigate a disease outbreak at Farm SI on March 13th. 
The goal is to identify farms at risk of infection, so she can develop effective strategies for disease control.
As disease outbreaks can propagate through complex interactions among farms, Jane uses SpreadLine to create a representation of the disease outbreak by providing essential information described in \autoref{sec:scenario}.
As shown in \autoref{fig:animal}-A, three timepoints of interest catch Jane's attention -- (1) the only animal senders before the outbreak; (2) unhealthy farms right before the outbreak; and (3) unhealthy receiving farms after the outbreak.
She prioritizes analyzing the latter two periods due to their health conditions. Based on the outbreak date, she infers that timepoint (2) could affect this outbreak and that the outbreak could further affect timepoint (3).
To better understand the farms in these two periods, she expands the blocks to examine their trading situation and geographical proximity (\autoref{fig:animal}-B)\revise{, in which the farms are repositioned in the expanded view based on their geographical locations.}
\revise{As shown in \autoref{fig:animal}-5, if Farm FCT affects the outbreak, its long visual distance to the ego (i.e., the larger point), indicating far geographical distance, suggests that the impact is likely through animal transportation.}
\revise{From \autoref{fig:animal}-6, Jane observes that while Farm FI exhibits better health conditions, its close visual proximity, suggesting short geographical distance, could contribute to outbreak impacts.}
By overlaying Farm FAO and FAN in other views, as annotated in \autoref{fig:animal}-7 and -8, Jane learns that the outbreak may consequently affect the two farms through transportation, as they are not within the ego's \revise{visual} proximity.


{\color{teal}\textbf{Function.}}
The straightened ego line, acting as a visual anchor, inherently divides the space into dichotomous compartments. 
We take this opportunity to distinguish important alters based on their directional relationships with the ego (\functionBox{T9}). 
In the scenarios illustrated above, we assign 1-level alters that transported animals to the ego to be positioned above the ego line, whereas those receiving animals from the ego are situated below. 
This assignment facilitates visual tracking of an entity when its relationship with the ego changes, e.g., the transition of Farm FCO from a sender to a receiver of animals in \autoref{fig:animal}-1.


{\color{Structure}\textbf{Structure.}}
Inspired by the station-style aesthetics of metro maps \cite{wu2020survey}, SpreadLine integrates block distinction to (1) indicate the entity presence in the egocentric network (\structureBox{T5}) and (2) accentuate entities of different identities (\structureBox{T4}). 
SpreadLine distinguishes 1-level and 2-level alters (structural identity) through different assignments to blocks --- 1-level alters are always positioned in the primary block with the ego, while 2-level alters stay in the secondary blocks, interconnected by slender edges. 
All 2-level alters are positioned based on the assignment of their connected 1-level alters, disregarding relationship direction (e.g., nodes C, D, and E at time 1 in \autoref{fig:encodings}).
For bi-directional relationships, SpreadLine prioritizes either the direction with higher edge weight or the one yielding a more optimal layout when both hold equal significance.
Acknowledging this design favors certain ego-alter relationships over others, SpreadLine presents the detailed structural information in the auxiliary view (\structureBox{T6}), as shown in \autoref{fig:animal}-4.




{\color{Strength}\textbf{Strength.}}
With the block distinction aligned in parallel throughout the timeline, SpreadLine can highlight frequent ego-alter interactions over time (\strengthBox{T8}).
Within the primary block, SpreadLines reorders the 1-level alters based on their edge weights, thereby prioritizing alters of higher relative importance (\strengthBox{T7}).
For example, at time 2 in \autoref{fig:encodings}, node D is positioned closer to the ego than node A is, due to its higher edge weight, despite both being 1-level alter.
SpreadLine does not apply the same design to two-level alters by default to maintain computational efficiency and layout quality during layout generation.

{\color{Content}\textbf{Content.}}
While the essential information from the above network aspects is woven into the layout generation, the content aspect is manifested through explicit visual encodings to contextualize the visual exploration of the egocentric network.
SpreadLine utilizes line colors to signify entity identities (e.g., production roles of farms in \autoref{fig:animal}) and node colors to reflect the ever-changing status of entities, e.g., health conditions (\contentBox{T11}).
We enhance the contextual understanding of the network by integrating a contextual affinity view at each timepoint (\contentBox{T10} and \contentBox{T12}).
\revise{This view repositions alters to reflect certain contextual relationships (e.g., geographical locations in \autoref{fig:animal}) and highlights the ego point in larger size. }
The contextual affinity view caters to more in-depth exploration, as our design objective is to ensure the user has first identified a specific timepoint or entity of interest.




\vspace{-0.05in}
\subsection{Layout Generation}

SpreadLine builds upon a foundational framework, StoryFlow~\cite{liu2013storyflow}, for its capacity to facilitate real-time layout generation.
However, we impose certain adaptations and additional constraints upon the framework to embody our unique design rationales.
As StoryFlow consists of three optimization stages: ordering, alignment, and compact, we proceed to illustrate the differences introduced to each stage.

\textbf{Ordering.} 
This stage relies on barycenter sorting to minimize line crossings of entities. SpreadLine first orders the entities based on their weights in the initialization and only performs the sorting technique on specific entity groups (i.e., 1-level alters of the same weight or 2-level alters), satisfying the design rationales for \function{} and \strength{}.

\textbf{Alignment.} 
StoryFlow aims to minimize line wiggles by formulating the comparison of two consecutive timestamps as finding the longest common substring with a reward mechanism. To ensure the ego line is consistently straightened, SpreadLine assigns the reward of self-alignment for the ego to be the maximum integer value.

\textbf{Compact.} 
StoryFlow employs integer linear programming to optimize both wiggle distance and white space utilization. 
For SpreadLine, as the relative importance of entities (edge weight) is encoded in their distance to the ego, wiggle distance optimization is incompatible with our objectives. 
SpreadLine needs to ensure that the lines of entities do not cross the ego line if they are to remain within the same compartment.
In comparison, SpreadLine adopts greedy heuristics that provide two distinctive optimization focuses: \textit{minimizing vertical spaces} and \textit{achieving more straight lines}. 
We refer to them as \textbf{vertical space optimization} and \textbf{straight line optimization} in the remaining manuscript.

\revise{\textbf{Vertical space optimization}} ensures the height of the block stays compact, i.e., entities maintain their minimal distances, in which lines of entities have to circumvent the blocks if they are absent in the network at a timepoint (referred to as ``idle lines'').
However, these design rationales introduce a tradeoff -- increased line crossings and wiggles -- imposing more visual clutter and reducing interpretability.
In response, SpreadLine offers another focus, minimizing line wiggles (i.e., more straight lines), at the expense of increased vertical white space. 
\revise{\textbf{Straight line optimization} attempts to align an entity to the same height across timestamps and} allows idle lines to traverse the blocks, where we signify their absence in the network by not positioning any points.
Although this optimization complicates the accomplishment of \structureBox{T5}, it enhances the visual tracking of entities.
\revise{The algorithmic logic and an example of both optimizations are in supplementary materials.}

\vspace{-2pt}
\subsection{User Interactivity}
As the primary objective of SpreadLine is to facilitate user comprehension of relationship dynamics among entities, we recognize that effective visual tracking of entities is crucial for successful visual exploration.
However, the efficacy of visual tracking diminishes as the number of entities or timestamps increases.
Consequently, SpreadLine supports a range of fundamental user interactions. 
\revise{\textbf{Horizontal scroll} allows users to navigate through timepoints in SpreadLine.}
\revise{\textbf{Hovering} over an entity line allows users to de-highlight others and focus on how the hovered alter interacts with the ego, thereby understanding the dynamics.}
\textbf{Clicking and pinning} provide users with a focused comparison of multiple entities of interest.
\textbf{Filtering} assists users in locating entities \revise{with distinct behaviors, such as longer presence in the network or important identity changes, facilitated by longer lines and the line crossings over the ego.}
\textbf{Tooltips} communicate the status of an entity at each timepoint, and timestamp \textbf{annotations} provide important temporal information (e.g., ``outbreak'' in \autoref{fig:animal}).


\subsection{Design Considerations}
\revise{Here, we outline major design considerations. Further discussions on line continuity and space division are in our supplementary materials.}

\textbf{Optimization focus.}
One aspect of storyline visualization research is layout optimization, driven by three metrics: white spaces, line wiggles, and line crossings.
SpreadLine offers two optimization focuses that minimize white spaces or line wiggles. 
While line crossings can cause visual clutter, we recognize that crossings over the ego line can indicate entity identity changes.
As these visual cues may provide useful insights, SpreadLine does not prioritize optimizing line crossings.

\textbf{Alter-alter communication.}
In the initial exploration of visual encodings in SpreadLine, we attempted to encode the alter-alter communication with triangular marks, motivated by their association with directional movements in semiotics \cite{borgo2013glyph}. 
However, this design introduces ambiguity, as we struggled to find an intuitive way to denote the receiving end of the directional relationships.
In further evaluation, we recognize the understanding of alter-alter communication can be deferred to a detailed exploration, i.e., the contextual affinity view.
We retain triangular marks within SpreadLine as a visual cue to signify entities' first and last appearances across the network.



\textbf{Centering.}
During the development of SpreadLine, we debated centering the ego in the contextual affinity view, similar to radial layouts in node-link representations.
Centralizing the ego, along with the ego line, ensures consistent attention to the ego, thereby facilitating the comparison of entities (\contentBox{T10}) due to the emphasis on relativity.
However, this design may prove less suitable when the trajectory of the ego in the contextual affinity view holds significance. 
It also results in a loss of contextual overview of the network.
Consequently, SpreadLine does not support this configuration in this prototype.

\textbf{Visual aggregation.}
Inherited from storyline visualization, SpreadLine faces similar challenges in visual scalability.
As entities populate block distinctions, the expanding vertical space amplifies the presence of line wiggles and line crossings, degrading visual clarity.
We considered visually aggregating neighboring entities to address this issue. 
Yet, as \strength{} is translated into the entity ordering, we have recognized that the many entities of equal importance will be aggregated. 
It may deviate from the general expectation of grouping entities with similar temporal behavior.
Unraveling the aggregation groups may further complicate user interactivity, as clicking would serve more functions beyond pinning the entities and expanding the blocks.
As SpreadLine has employed various visual encodings to translate diverse aspects of network information, we have opted not to introduce visual aggregation in the current prototype to maintain simplicity.
Instead, SpreadLine offers the functionality to \textbf{stack the entities} based on their contextual identity (line color), as demonstrated in \autoref{sec:case-coauthors}.

\section{Using SpreadLine}
\revise{
Using SpreadLine concerns two user groups --- framework users and visualization consumers.
}
In this section, we focus on the practical utilization of SpreadLine \revise{for framework users}.
SpreadLine is designed for designers seeking a comprehensive understanding of data through visual exploration. 
Our framework caters to a diverse user base, including but not limited to, data collectors, domain experts, and data analysts.
Our focus begins with illustrating how users can customize SpreadLine to align with their diverse exploration needs, followed by how users can communicate with the SpreadLine framework programmatically.

\begin{figure*}[h]
    \centering
    \includegraphics[width=0.98\linewidth,]{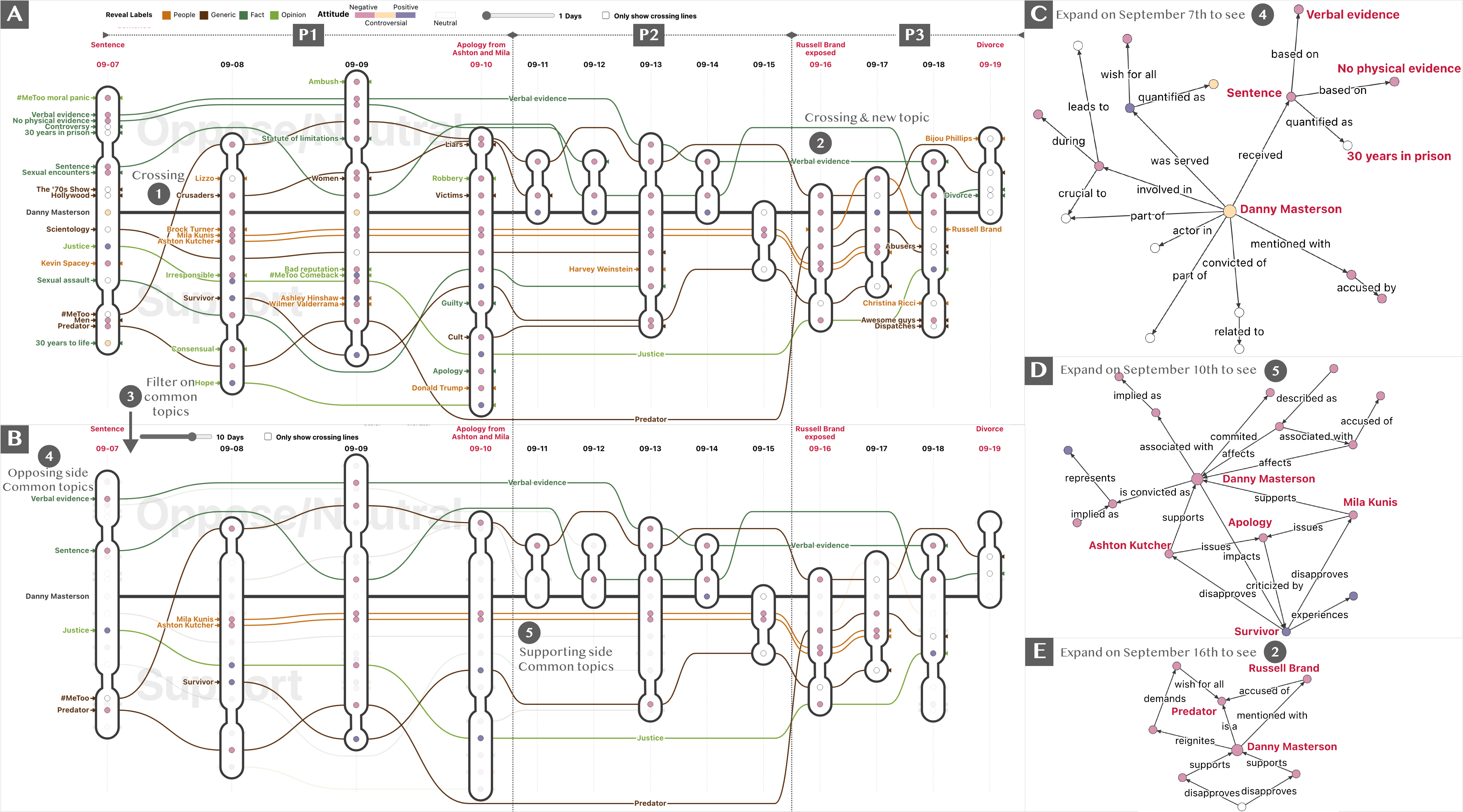}
    \caption{Public reaction to the sentence of the actor Danny Masterson. (A) The overview. Based on the public engagement, reflected in block distinction, we identify three distinct periods (P1, P2, and P3). (B) The focused examination on common topics. The slider is repositioned for presentation. The following are knowledge graphs of tweets on different days (entities are annotated for presentation): (C) On September 7th, opposing sides discussed evidence associated with the sentence. (D) On September 10th, the supporting sides condemned the two supporters of the ego and refused their apology. (E) On September 16th, there was a resurgence of public reaction due to the involvement of another individual in the \#MeToo movement. \revise{Note that \textbf{vertical space optimization} is employed with entity stacking.}}
    \label{fig:case-metoo}
    \vspace{-0.15in}
\end{figure*}

\subsection{Customization Guidelines}
As discussed in \autoref{sec:egocentricl-visual-analysis}, the visual exploration of intricate egocentric networks has to accommodate various analytical requirements. 
Acknowledging the pivotal role of data diversity in this context, we design SpreadLine to address users' requirements through its adaptable configurations.
In the preceding section, we have introduced a foundational configuration; here, we provide insights into configuring SpreadLine to harness its inherent flexibility.
The set of encodings available for customization is outlined in \autoref{tab:encoding_rec}.

Our primary objective behind space division and block distinction is the effective characterization of entities. 
\textbf{Space division} leverages spatial regions, while \textbf{block distinction} exploits shapes; both are effective identity channels for categorical attributes \cite{munzner2014visualization}.
We recommend designers employ these two encodings to furnish users with a holistic understanding of alters. 
For instance, \revise{transaction roles and 1-level and 2-level alters portrayed in \autoref{fig:animal}} provide domain experts with an overview of disease propagation through the trade network.
The crossings in space division or block distinction carry varying levels of significance: while both denote entity identity changes, crossings in space division involve the ego line. 
\revise{The crossings over the ego line can serve as a visual cue to highlight significant changes in the behavior of alters. Therefore, we recommend using contextual information for space division to enrich the overall data exploration process.}

\textbf{Line color} enables the encoding of an additional identity of alters, for which we recommend using contextual categorical information to highlight group patterns.
For example, in \autoref{sec:case-coauthors}, we use line color to indicate the initial role of a co-author to the ego, revealing shifts from internal to external collaborators (and vice versa).
%
\textbf{Line ordering} uses magnitude channel (position on an unaligned scale \cite{munzner2014visualization}) to emphasize entities more important to the ego, such as more animals in transport may imply a higher disease propagation rate due to close contacts.
Similar to \textbf{node color}, both encodings aim to guide users toward identifying entities of interest; line ordering is influenced by edge weight (strength), while node color considers contextual information (content).
Given that the network topology reveals structural relationships among entities, our intention for the \textbf{contextual affinity view layout} is to focus on their contextual relationships.

As the scale of data dimensions increases, so does the design space for SpreadLine.
Through the guidelines, we aim to assist designers in navigating their design spaces by sharing our design intentions for these customizable encodings.
It is important to note, however, that these guidelines do not represent rigid boundaries for the design space, nor do they suggest that the utilization of SpreadLine requires the data to include all four network aspects.
Depending on the specific application scenarios, we believe there are alternative configurations that may offer more effective avenues for data exploration. 
For instance, network analysts
may find graph metrics such as connectivity or transitivity to be more instrumental in deducing relationships between entities. Consequently, this information can be incorporated into the contextual affinity view, despite being derived from the network topology, to provide more context to domain experts.
By aligning encodings with data properties and designer intentions, our objective with these guidelines is to empower potential users of SpreadLine to leverage its flexibility.

\begin{table}[t]
    \centering
          \resizebox{2.7in}{!}{ 
    \begin{tabular}{p{0.25\linewidth}  p{0.28\linewidth}  p{0.3\linewidth}}
        \toprule
         \textbf{Encodings} & \textbf{Recommended configuration} & \textbf{Primary objective}  \\
         \midrule

         Space division& 
         Function or contextual identity (1D) & High-level characterization of entities \\
         \arrayrulecolor{black!30}\midrule

         Block distinction&
         Function or structural identity (1D) & Low-level characterization of entities \\
         \arrayrulecolor{black!30}\midrule

         Line color&  
         Entity contextual information (1D) & Highlight patterns of entity groups \\
         \arrayrulecolor{black!30}\midrule

         Line ordering&
         Strength (1D) & Relative importance to the ego \\
         \arrayrulecolor{black!30}\midrule

         Node color&
         Entity contextual information (1D) & Highlight entities of interest \\
         \arrayrulecolor{black!30}\midrule

         Contextual affinity layout&
         Multivariate contextual information & Relationships among entities \\
         \arrayrulecolor{black}\bottomrule
         
    \end{tabular}
    }
    \caption{Customization offered in SpreadLine. We include the foundational configuration as recommendation and describe the designer intentions behind these options. 1D refers to univariate information.}
    \label{tab:encoding_rec}
    \vspace{-0.2in}
\end{table}

\subsection{Framework Interface}
In this subsection, we focus on the interface for the target users to specify their customizations in SpreadLine. \revise{Our source code and framework interface can be found in supplementary materials.}

\textbf{Implementation.}
There are two components in the implementation of SpreadLine: layout generation and rendering.
SpreadLine operates by taking a predefined schema and the data as input. 
Users must specify different aspects of the data in the schema for SpreadLine to interpret the contained information, e.g., mapping "farm name" to "entity".

\textbf{Data requirements.}
The minimum data requirement in SpreadLine includes the temporal relations of entities and the specification of the ego.
By default, SpreadLine captures a 2-level egocentric network for the defined ego.
However, to fully exploit the capabilities of SpreadLine, a minimum of four data dimensions is necessary to contextualize the data exploration, particularly if the relations already contain edge types and edge weights. 
For instance, in the context of the animal disease outbreak, we include farm production roles (line color), health conditions of farms (node color), and the geospatial locations of farms (contextual affinity layout).

\textbf{Hyperparameters.}
SpreadLine offers various customizations through hyperparameters, including (1) choosing between optimization focuses (straight line or vertical space); (2) whether to stack the entities based on their contextual identity (line color); (3) stretching the distance between certain timestamps to mitigate the visual clutter caused by line crossings; (4) whether to present the entity relations through a node-link representation in the contextual affinity view; and (5) timestamp annotations to enrich the temporal context.



\section{Evaluation}
\label{sec:evaluation}
We have described how a domain expert user could utilize SpreadLine to understand the influence of a disease outbreak among animal farms.
To illustrate the generalizability of SpreadLine in various domains, 
we present two more case studies on different real-world datasets. 
For datasets, we collect social media posts to show how we construct networks from unstructured texts to analyze the trends of topics and we use an academic publication dataset to exemplify social network construction from event sequence data.
We also conduct a usability study to evaluate the interpretability and effectiveness of SpreadLine. 
More details can be found in supplementary materials.

\begin{figure*}[h]
    \centering
    \includegraphics[width=0.96\linewidth,]{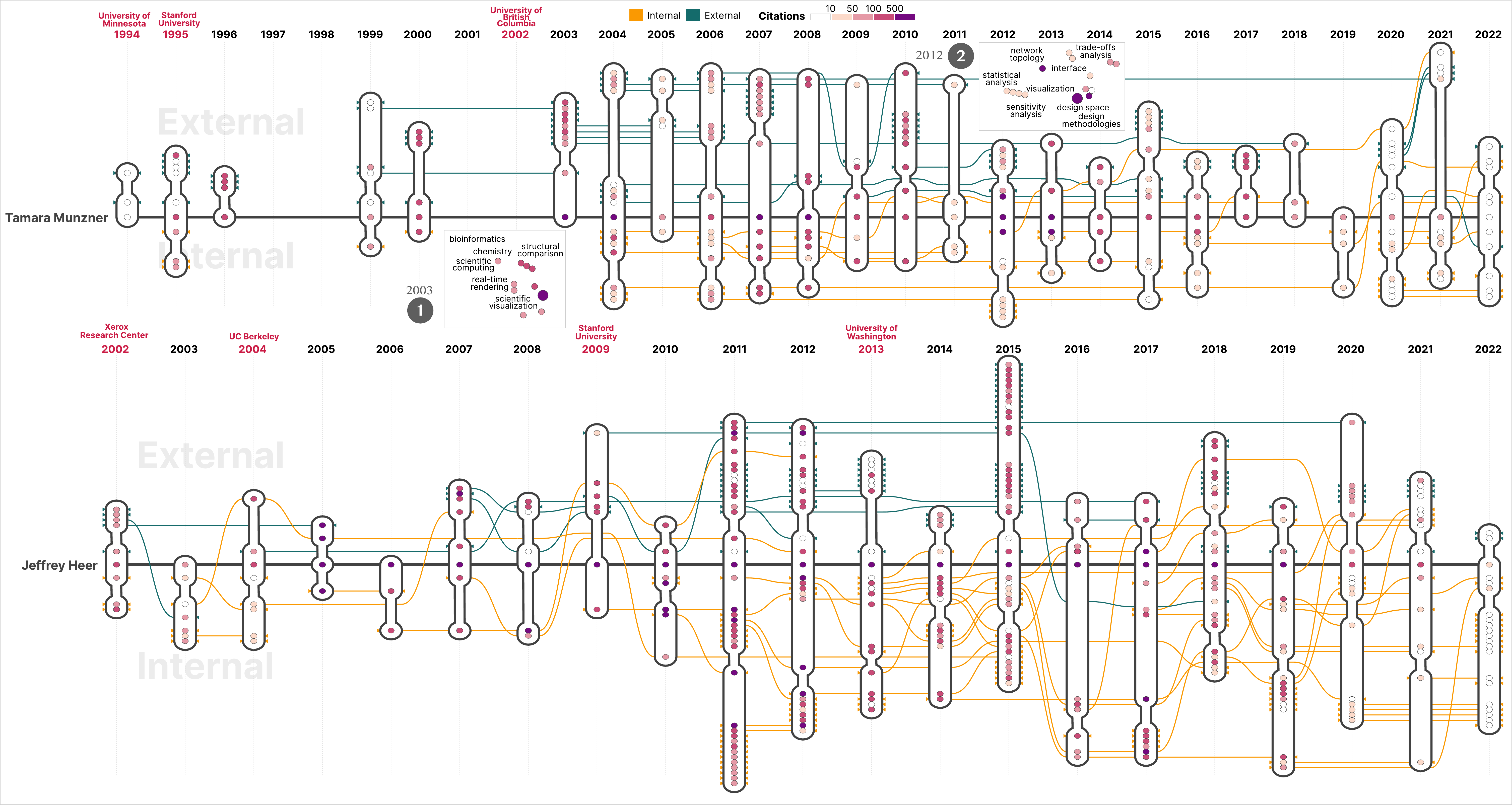}
    \caption{Career evolution of two researchers. Top: Tamara Munzner with 123 co-authors from \revise{\textbf{1994}} to 2022. Bottom: Jeffrey Heer with 198 co-authors from \revise{\textbf{2002}} to 2022. (1)(2) Research focus of Tamara in 2003 and 2012, respectively. \revise{\textbf{Straight line optimization} is employed with entity stacking.}}
    \label{fig:case-coauthor}
    \vspace{-0.15in}
\end{figure*}

\subsection{Case 1: Public Reaction to \#MeToo Event}
\label{sec:case-metoo}
In September 2023, American actor Danny Masterson was sentenced to 30 years to life in prison, becoming the first major Hollywood figure convicted of sexual assault in the \#MeToo movement.
To understand the public reaction to this trial, we collect 48 relevant posts on the social media platform X (formerly Twitter) from September 7th to September 19th.
We consider a post relevant if there is both the mention of Danny Masterson and the \#MeToo hashtag.
We are interested in understanding: (1) The public's stance for the sentence of Danny Masterson (support, oppose, or neutral) and (2) The public's attitude towards related topics (positive, negative, controversial, or neutral). Consequently, we designate Danny Masterson as the ego.

\textbf{Data processing.} 
We consider knowledge graphs \cite{peng2023knowledge} robust for extracting the essential concepts from unstructured texts.
In particular, large language models are shown to be more reliable in information extraction \cite{li2023evaluating} and annotation tasks \cite{tornberg2023chatgpt, gilardi2023chatgpt}; therefore, we employ ChatGPT-4 \cite{openai2024chatgpt} to perform the following tasks: (1) annotate the {\color{Content}\textbf{stance}} of each post for the sentencing of the ego ; (2) construct a {\color{Structure} \textbf{knowledge graph}} for each post; (3) determine the poster's {\color{Content}\textbf{attitude}} towards each entity in a graph (positive, negative, or neutral); and (4) {\color{Content}\textbf{annotate}} whether each entity is objective fact (if not, subjective opinion), or explicit mention of individuals (if not, generic groups).
For task (3), at each timestamp for each entity, we compare the ratio of positive tags over negative tags to further determine whether an entity is rather controversial.
For task (4), we ask ChatGPT-4 to annotate based on its knowledge regarding the event, where no posts are provided.
This process yields a collection of 49 entities associated with 178 relations.


\textbf{Encoding design.} 
Given our primary objective is to understand the public reception of the ego, we choose \textbf{space division} for the post standings.
\textbf{Line color} characterizes the properties of entities derived from task (4), whereas \textbf{node color} addresses public reception towards specific entities (\contentBox{T11}).
\revise{We use 1-level and 2-level alters for \textbf{block distinction}}, as 2-level alters in ego's knowledge graphs can offer additional context (\structureBox{T4}). 
\textbf{Line ordering} considers the number of mentions at each timestamp (\strengthBox{T7}).
In the \textbf{contextual affinity view}, we present the entire knowledge graph at each timestamp using a force-directed layout, since the annotated relations will provide complete context for users to understand the relationships between these entities (\structureBox{T6}, \functionBox{T9}).
This case stacks entities and employs vertical space optimization.

\textbf{General trends.}
We focus on the global pattern and a set of findings we have found through the filtering interaction supported by SpreadLine.
Based on the network population (\structureBox{T1}, \structureBox{T5}) that reflects public engagement, as shown in \autoref{fig:case-metoo}-A, we identify three distinct periods: (P1) The initial three days following the sentence (September 7th to 10th); (P2) A relatively calm period (September 11th to 15th); and (P3) A resurgence of public reaction (since September 16th).
Throughout the timeline, the reception of the ego remains controversial or fluctuates between positive and negative criticism. 
Generally, both sides seem to hold negative attitudes toward most topics.
As shown in \autoref{fig:case-metoo}-1 and 2, given the few crossings over the ego line, we may infer that both sides pay attention to different topics (\contentBox{T11}).

\textbf{Topic context.}
We further explore the common topics discussed on each side by utilizing the \textbf{filter} on entity lifespan (\strengthBox{T8}), as shown in \autoref{fig:case-metoo}-B. 
When we focus on topics that have been discussed for a minimum of 10 days, we observe the following: (1) Most topics in P2 began in P1; (2) The opposing side tends to discuss \textit{sentence}, \textit{verbal evidence} (both are annotated as facts), and \textit{\#MeToo}; and (3) The supporting side concentrates on two individuals (\textit{Ashton Kutcher} and \textit{Mila Kunis}), \textit{Justice}, \textit{Survivor}, and \textit{Predator}.
We can gain a deeper understanding of these topics by \textbf{clicking} on the blocks involving these entities. 
For example, on September 7th, the opposing side discussed that the life \textit{sentence} of the ego is based on \textit{verbal evidence}, as shown in \autoref{fig:case-metoo}-C.
On September 10th (\autoref{fig:case-metoo}-D), we observe that the two individuals have shown support for the ego.
We find \textit{survivor} disapprove of these two supporters and criticize their apology.
One may follow the same process to learn about other topics (\structureBox{T6}, \functionBox{T9}).

\textbf{Period differences.}
We may conclude the main difference between P1 and P2 is public engagement as they share many topics (\structureBox{T2}).
While it may be reasonable to assume that the level of public engagement likely decreases over time in the aftermath of an event, further investigation is required to understand the dynamics in P3.
Consequently, as shown in \autoref{fig:case-metoo}-2, we look for any emerging topics around P3 and identify a topic of interest, \textit{Russell Brand} (\structureBox{T3}).
Following the same process, as revealed in \autoref{fig:case-metoo}-E, we find that the ego is mentioned with this individual, where both are accused of being \textit{predators}.
This finding helps us conclude that the ego regained public attention in P3 due to a newer predator accusation against another individual (\contentBox{T13}).


\subsection{Case 2: Career Evolution of VIS Researcher}
\label{sec:case-coauthors}
Existing works \cite{fu2021dyegovis, peng2018dmnevis} have constructed collaboration networks from research papers to explore the social circles of researchers.
We also utilize a popular DBLP publication dataset \cite{tang2008arnetminer} to analyze researchers in the visualization field.
As this information inherently represents a researcher's devotion to their career, we are rather interested in exploring (1) the collaboration patterns of a researcher and (2) the research focus change over time. This leads to a different network construction.


\textbf{Data processing.}
We only consider the following information in this dataset: (1) Author names in order; (2) Author affiliations in order; (3) Published year; (4) Citation count up to January 2023; and (5) Author-defined keywords.
We use {\color{Structure} \textbf{the author order}} to construct the network for each paper.
The {\color{teal}\textbf{first author}} always has links going outwards to other non-first co-authors.
This suggests that if a researcher's node outdegree is 0, they have only served as non-first co-authors; otherwise, they have been first authors at least once. 
Once the ego is specified, we annotate the alters as internal collaborators if they are from the same {\color{Content}\textbf{affiliation}}; otherwise, external.
We consider author-defined keywords to represent the researcher's focus at the time.
Consequently, we employ ChatGPT-3.5 to extract the word embeddings for each keyword \cite{raval2023explain} and apply dimensionality reduction \cite{van2008visualizing} to obtain a 2-dimensional representation of conceptual similarity among keywords across the timeline.
Each researcher is assigned representations as their research profile based on the papers they worked on at each timestamp.

\textbf{Encoding design.}
As our exploration focus is collaboration tendency, we choose \textbf{space division} for the affiliation identity (internal or external), whereas \textbf{line color} records their identity when they first join the egocentric network (\contentBox{T11}).
For \textbf{block distinction}, we separate first and non-first co-authorships with node outdegrees (\structureBox{T4}).
\textbf{Node color} represents the total citation that a researcher gained from the collaboration with the ego, and \textbf{line ordering} encodes the total number of publications from the ego-alter collaboration across time (\strengthBox{T8}). 
In the \textbf{contextual affinity view}, we use the research profiles to visualize the similarity of research focus and annotate the keywords to support the interpretation (\contentBox{T12}). 
We annotate when ego changes their affiliations.

\textbf{General trends.}
We first focus on Tamara Munzner's career evolution, as shown in \autoref{fig:case-coauthor}-top.
We may consider the evolution as three periods: (1) Before 2003; (2) 2003 -- 2011; and (3) 2012 to the present.
The first two periods are considered different due to the network population, whereas we use the citation trend to tell the last two apart.
In each period, we may examine the research focus in the years with high citations to gain more context. For example, as annotated in \autoref{fig:case-coauthor}-1 and 2, we observe that Tamara worked more on scientific visualization in 2003 and then moved on to visualization design space in 2012.
We find she has a rather balanced external and internal collaboration.


\textbf{Visual comparison.}
We also examine Jeffrey Heer's career evolution, as shown in \autoref{fig:case-coauthor}-bottom and compare it to Tamara's.
Based on the network population, we may consider his career to have two periods, anchored by the year 2011. 
In many years of his career, he received more than 500 citations every year.
One may further break down the first period from 2005, the first impactful year.
When compared to Tamara, our main finding is that Jeffrey has more collaborators, particularly internal, in his 12-year career than Tamara has.
When both started their career, Tamara had more external collaboration, but Jeffrey had a balanced mixture.
While we find both had internal collaborators becoming external, due to the line crossings over the ego line, they exhibit different patterns. 
As Tamara has not changed her affiliation since 2002, we conclude that it is her collaborators leaving the affiliation.
Jeffrey had the same pattern since 2013, but we find the crossings before 2013 mostly correspond to his affiliation change instead.


\textbf{Participants insights.}
\revise{During the usability study,} all of our participants were able to derive insights mentioned above, e.g., active periods and collaboration tendencies.
Some compared the collaboration tendencies across individuals: Tamara has relatively more external collaboration.
Many participants pointed out a commonality between these two researchers: Their careers likely took off after around 10 years, and then they started expanding external collaboration, but in recent years, both turned to internal collaboration more.
For their differences, one participant mentioned Jeffrey tended to produce impactful papers more with his internal first co-authors, which is different from Tamara. 
After exploring both researchers, two participants with experience in visualization suggested that Tamara seemed to focus on very different topics over time, which may explain her shorter collaboration, whereas Jeffrey seemed to work a set of core problems over time, which could be the reason for more of his long-term collaboration.


\subsection{Usability Study}
\label{sec:usability}
We utilize the dataset introduced in \autoref{sec:case-coauthors} to conduct \revise{an in-person} usability study. 
Our goals are to evaluate: 
(G1) the capability of SpreadLine supporting the egocentric network analysis tasks introduced in \autoref{fig:taxonomy}
and (G2) the utility and usability assessment of SpreadLine.

\textbf{Participants.}
We recruited 10 participants (5 males and 5 females; aged 18--34) with different levels of experience in visualization (visual analytics, network visualization, and computer graphics) \revise{from a local university}. 
4 of them have 3 -- 6 years of experience in visualization, referred to as \textbf{experts}; 3 of them have 0 -- 2 years of experience instead (\textbf{beginners}); 3 of them have no experience with visualization (\textbf{novice}).
We recruited novices as \revise{they are potential visualization consumers of SpreadLine}.
While all experts and beginners know about network visualizations and academic publication datasets, none were familiar with this specific dataset.
All novice participants were unfamiliar with both topics mentioned above.
All the participants had normal or corrected-to-normal vision without color vision deficiency\revise{, and none had prior knowledge of SpreadLine}.

\textbf{Task design.}
For G1, we developed 11 quantitative tasks to evaluate participant responses regarding correctness and time. 
3 qualitative tasks are designed to observe how participants obtain insights from SpreadLine. 
Each prompt corresponds to a task in the taxonomy:

\begin{itemize}[label={}, noitemsep, leftmargin=*]
    \item \structureBox{T1} Report any insights you find in the visual representation.
    \item \structureBox{T2} Compare two egos and report any commonalities or differences.
    \item \structureBox{T3} When is the first (last) appearance of \textit{alter} in the network?
    \item \structureBox{T4} How does \textit{alter}'s relationship change with ego?
    \item \structureBox{T5} Which year has the most population?
    \item \strengthBox{T7} Who published relatively more papers with ego?
    \item \strengthBox{T8} Who are long-term collaborators with ego over certain years?
    \item \functionBox{T9} How many times has \textit{alter} been a non-first author with ego?
    \item \contentBox{T10} Which first co-authors publish more impactful papers with ego? 
    \item \contentBox{T11} In what years does the ego have more than 500 citations?
    \item \contentBox{T12} How did the ego's research focus change over specific years?
    \item \contentBox{T13} Summarize ego's career evolution.
\end{itemize}

Most quantitative tasks are freeform questions, except for two multiple-choices (\structureBox{T4} and \strengthBox{T7}), with at least four choices, and a ranking question \contentBox{T12}.
Participants can use \textbf{filtering} and \textbf{pinning} besides highlights in SpreadLine to answer these questions.
We generate static images for the qualitative questions, \structureBox{T1}, \structureBox{T2}, and \contentBox{T13}. 
We ask participants to answer \structureBox{T1} for SpreadLine and \textbf{node-link} representations. The images can be found in the supplementary materials.



We did not evaluate \structureBox{T6} in this study as it is incompatible with our goals of exploring academic collaboration networks.
For simplicity, we only evaluate \textbf{straight line optimization} of SpreadLine with the expectation of increased difficulty for completing \structureBox{T5}.
A focused mode is added in this study. 
We only annotate the top 20\% entities with the longest lifespan in the network, at a maximum of 25 labels.
We inform participants that all the alters specified in the tasks can be found in this mode. 
Participants will only answer one question for one alter.


\begin{table}[t]
    \centering
      \resizebox{2.00in}{!}{ 
    \begin{tabular}{c|c|c}
        \toprule
        \textbf{Task} & \textbf{Task Accuracy} & \textbf{Task Time} (secs) \\ 
        \arrayrulecolor{black!30}\midrule

         \structureBox{T3} & 100\% & 27.25 (13.36)\\
         
         \structureBox{T4} & 98\% & 12.97 (2.36) \\
         
         \structureBox{T5} & 100\% & 83.53 (31.19) \\

         \strengthBox{T7} & 95\% & 64.42 (30.23) \\

         \strengthBox{T8} & 99.2\% & 23.03 (8.84) \\

         \functionBox{T9} & 90\% & 38.70 (10.52)\\

         \contentBox{T10} & 100\% & 48.73 (18.68)\\

         \contentBox{T11} & 100\% & 20.33 (7.93)\\

         \contentBox{T12} & 97.5\% & 69.11 (20.59)\\
        \bottomrule
    \end{tabular}
    }
    \caption{Completion times and accuracy for each task. Task time is shown in average (standard deviation) in seconds.}
    \label{tab:evaluation-performance}
    \vspace{-0.3in}
\end{table}

\textbf{Procedure.}
The study begins with a brief introduction to the dataset and SpreadLine. 
We adopt a similar approach to \autoref{fig:encodings} to explain the network construction and the encoding design of SpreadLine.
Participants then watched a clip to familiarize themselves with the available user interactions.
We prepare a training session for participants to get familiar with SpreadLine by showing a different researcher.
This researcher is on a smaller data scale
and has had only one collaboration in total with the two researchers in the actual experiment.
Participants were prompted for the correct answers at the end of the training session.

In the testing session, for the first researcher, participants went through the quantitative tasks in a randomized order, then performed \structureBox{T1} with different visual representations (the order is counter-balanced), followed by \contentBox{T13}.
For the second researcher, participants only answered \structureBox{T4}, \strengthBox{T8}, \functionBox{T9}, \contentBox{T12}, and \contentBox{T13}. 
The session ends with \structureBox{T2} by comparing these two researchers.
There was no time limit and participants had the option of skipping the questions. 
Task completion times and participant answers are recorded.
We conducted observations and screen-captured the entire session.
After the study, participants were asked to provide an assessment of SpreadLine. The study lasted around 1.5 hours and was conducted on a MacBookPro (2.4 GHz Quad-Core Intel Core i5, 16GB 2,133 MHz LPDDR3), connected to a 27-inch display (1,920 $\times$ 1,200 pixels) with the Chrome browser.

\textbf{Results of G1.}
\autoref{tab:evaluation-performance} summarizes the results from our study. 
\textbf{
Most participants answered correctly using SpreadLine in a reasonable response time.
}
\structureBox{T4}, \strengthBox{T8}, and \contentBox{T11} exhibit rather consistent completion times across participants.
When asked about the most difficult question, three participants mentioned \strengthBox{T7}, two said \structureBox{T5}, and the other two chose \contentBox{T12}.
These three questions are also the ones with longest completion times and highest variations.
The reason why \strengthBox{T7} is difficult may vary. 
B1 attributed it to the scalability issue, \textit{``It is harder to tell as the network population grows.''}.
E3 considered line ordering rather implicit when compared to other encodings, with which B3's comment resonates that it is harder to recall this encoding. 
%
Both E1 and N3 pointed out \structureBox{T5} is harder due to the stretched vertical space, aligning with our expectation of the trade-off for optimization focuses. 
As for \contentBox{T12}, E2 attributed the cause to the consistent clicking on the blocks, and N1 mentioned their unfamiliarity with visualization.

\textbf{Participants find SpreadLine offers unique insights effectively through space division, line continuity, and block distinction}, in the comparison of SpreadLine and node-link representations (\structureBox{T1}).
For unique insights, 9 participants mentioned the organization change of co-authors, facilitated by \textbf{space division}. 
7 participants brought up \textbf{line continuity} for tracking entities and revealing long-term collaboration.
For insights in node-link diagrams, 4 participants mentioned alter-alter communication, while E1 commented \textit{``Although I may not need this information if my focus is the ego, which it is.''}.
For certain insights that both reveal, population change and distribution of non-first and first co-authors, 6 participants mentioned SpreadLine is more effective at showing the nuances thanks to \textbf{block distinction}.
As for \structureBox{T2} and \contentBox{T13}, unique participant findings are summarized in \autoref{sec:case-coauthors}.




\begin{figure}
    \centering
    \includegraphics[width=1\linewidth]{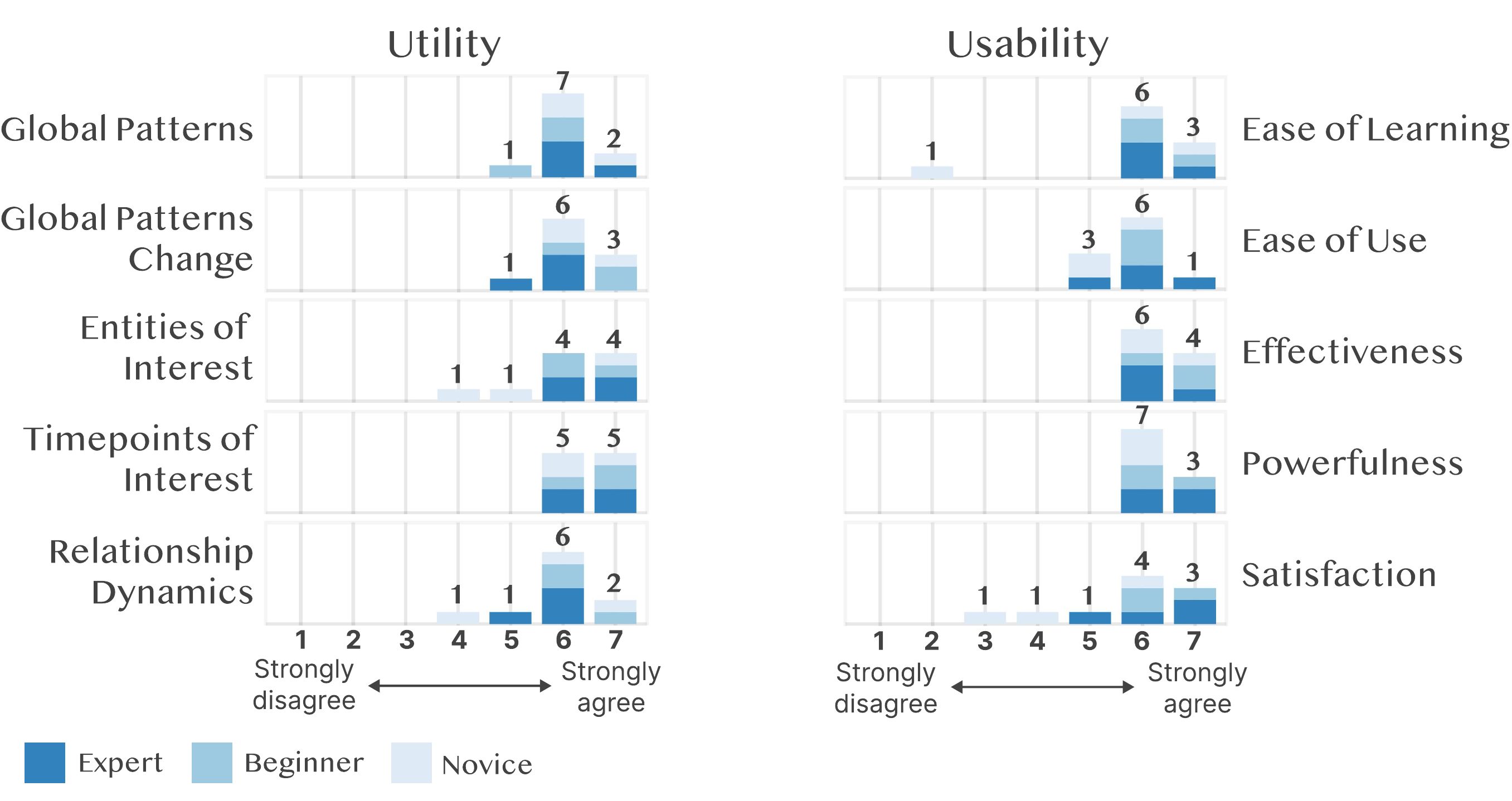}
    \caption{Histograms of participants' ratings (7-point Likert scale) on the overall utility and usability of SpreadLine.}
    \label{fig:evaluation-usability}
    \vspace{-0.2in}
\end{figure}

\textbf{Results of G2.} 
\autoref{fig:evaluation-usability} presents the participants' ratings on the overall utility and usability of SpreadLine. 
\textbf{SpreadLine generally is recognized for its capability, with an emphasis on timepoints of interest.}
Some participants comment that \textit{``SpreadLine offers a good way to communicate the temporal evolution''} (E2) and \textit{``You can easily tell if anything happens in a year through the line crossings or the blocks.''} (B1).
Participants also appreciate that SpreadLine helps them identify long-term collaborators (B1, E1, E3, and E4) or locate those with similar expertise as ego (E1, E3, and N3).
SpreadLine is also acknowledged for effectively showing changes in population, citation, and organization (B2, B3, E1, and N3), while more effort is required for research focus due to consistent clickings (E2 and E4).
\textbf{Visually connecting the four network aspects is considered a core strength of SpreadLine.} 
In the usability assessment, most participants appreciate SpreadLine for its effectiveness ---
\textit{``It shows different levels of information clearly and can highlight the evolving impact of the ego network for me.''} (E3)
and \textit{``Seeing everything connected and placed in context makes it much easier to understand what is going on.''} (E1).
\revise{However, we observe that the interpretability of SpreadLine varies among participants.}
All participants except N2 agree that SpreadLine is easy to learn. 
Struggling with qualitative questions, N2 explained, \textit{``I think SpreadLine is useful, but I need more time to familiarize myself more to conclude various insights SpreadLine offers.''}.
\revise{Some participants, like E1 and N3, found the design intuitive and understood it quickly, while E2 and N1 noted that even though they grasped the design during the tutorial, they could see others struggling. 
We recognize that SpreadLine can be challenging to consume without guidance; we plan to implement onboarding tools to support various visualization consumers in learning SpreadLine.}
Participants also suggest improvements to support \structure{} and \content{}.
For instance, E1 suggested using rules to indicate the population threshold to enhance the support of \structureBox{T5}, whereas N1 would like to adopt the word cloud styles in the contextual affinity view for easier summarization (\contentBox{T12}).

\vspace{-4pt}
\section{Discussion}
\revise{SpreadLine is generalizable so long as (1) the data describes the relationships of entities and can be represented as networks and (2) the ego can be specified (e.g., analysis target). Our case studies have also demonstrated three different contexts: (disease) propagation, (social media) trend, and (researcher) interactions.}
We further discuss the limitations and future potential of SpreadLine.

\revise{
\textbf{Scalability of SpreadLine.}
SpreadLine embeds essential network information in layout generation to effectively convey various insights.
This design presents a double-edged sword: It constrains the layout optimization, imposing visual clutter with more complex and larger datasets (e.g., more timestamps).
For example, line crossings are a primary contributor to visual clutter, although our study participants confirmed that crossings over the ego line provide unique and informative findings.
To mitigate visual clutter, SpreadLine provides filtering or scrolling interaction for visualization consumers and time interval configurations for framework users.
SpreadLine is also concerned with data scalability, the trade-off between visualized information and data quantity, as the same dataset can be constructed as different networks.
During the data processing in~\autoref{sec:case-metoo}, we employed different graph processing techniques and obtained networks of varying sizes. 
While larger networks can contain more complete information, they can also include uninformative entities due to lack of context (e.g., \textit{physical}). 
To manage the trade-off, we chose the network representation that balances context and relevance to our analysis needs.
Certain graph processing techniques, such as knowledge graphs \cite{chen2020review}, can capture essential concepts with fewer entities and relations in the networks, providing visualization consumers with information that is manageable and useful.
To handle larger networks, we plan to explore multi-level visual summaries for SpreadLine to better represent group patterns.
}


\textbf{Encoding design in SpreadLine.}
Most encodings in SpreadLine use space or color channels, leading to two competitions: (1) Line ordering and space division and (2) line color and node color.
In our study, space division is considered an effective encoding, while many participants found line ordering the least effective, limiting the support of \strengthBox{T7}.
One possible explanation is the conflicting design rationales of these two encodings.
Our intention for line ordering is to highlight important alters through a close alignment to the ego, akin to how story characters that often accompany the protagonist are considered significant by the audience.
We also consider this concept to align with the grouping cues through position proximity \cite{franconeri2021science, tversky2001spatial}, one of the more powerful techniques to highlight entity groups \cite{brooks2014traditional}. 
However, the space division relies on broader spaces to inform global patterns, i.e., the opposite visual cues, which may consequently limit the interpretability of line ordering.
On the other hand, color plays an important role in SpreadLine as it can serve exploration guidance effectively \cite{wolfe2017five}.
SpreadLine remains effective when the line color uses a categorical color scale (no more than 3 distinct colors) and the node color a sequential one. 
In the \#MeToo case study, we experimented with two categorical color scales but had to lower the saturation of one to mitigate the competition.
\revise{We have evaluated the analysis capability of SpreadLine and its usability for visualization consumers in \autoref{sec:usability}. Moving forward, we will develop a concrete design space for SpreadLine and investigate the practical utilization of SpreadLine for framework users.}


\textbf{SpreadLine as a visual data story authoring tool.}
During our study of qualitative questions, we confirmed that SpreadLine can offer entry points for participants to remain engaged in data exploration --- \textit{``Looking at the static images, it is already easy for me to point out several insights''} (E1), \textit{``I can keep exploring and reporting my insights if time is not an issue''} (E4), and 
\textit{``I think SpreadLine is strong for bringing in other people in the network exploration''} (B2). 
When asked about creating a visual data story with SpreadLine, several participants expressed their interest in the potential and suggested functionalities they would like to see in the authoring tool.
Besides animation support and annotation tools, two participants mentioned their preference for marking specific periods and drilling down to each period for a more in-depth exploration.
The development of such authoring tools for data storytellers \cite{shi2020calliope, morth2022scrollyvis} is thus desired, and it is promising
to leverage the concept of characters in creating a visual data story \cite{dasu2023character, ma2011scientific}.

\vspace{-4pt}
\section{Conclusion \revise{\& Future Work}}
We introduce SpreadLine, a visualization framework that leverages the storyline-based design to facilitate egocentric network exploration.
The development of SpreadLine is guided by a taxonomy of egocentric network analysis, characterized by four distinct network aspects.
SpreadLine offers customizable encodings with guidelines to meet diverse analytical requirements and exploration needs.
We demonstrate the applicability of SpreadLine in diverse real-world case studies.
The usability study \revise{of visualization consumers} confirmed the capability of SpreadLine to connect the four network aspects visually.

\revise{
Future work for SpreadLine includes (1) assisting visualization consumers in story comprehensions, such as onboarding tools or the detection of distinct entity behavior; (2) developing practical guidelines for framework users and evaluating how they design with SpreadLine; and (3) designing a data story authoring tool for SpreadLine.
}


\section*{Supplemental Materials}
All supplemental materials are available at \url{https://github.com/yunhsinkuo/SpreadLine}.
They include (1) source code, (2) a sample script of using SpreadLine and a demo web application, (3) optimization details, (4) additional design considerations, (5) datasets for two case studies and the associated large language model prompts, (6) two SpreadLine representations on a larger data scale, and (7) self-reported background information of our usability study participants.

\acknowledgments{%
This research is supported in part by the UC Climate Action Initiative and the National Institute of Health via grant R01CA270454, and in part by the National Science Foundation with grant no. ITE-2134901. %
}

\bibliographystyle{abbrv-doi-hyperref}
\balance
\bibliography{template}

\end{document}